\documentclass[prd,aps,preprint,showpacs]{revtex4}

\begin{document}
\def\slash#1{#1\!\!\!\!\!/}
\def\rpv{\slash{R_p}~}

\begin{titlepage}
\voffset 1.5cm

\preprint{KIAS-P03048, \hspace{1ex}
MC--TH--2003--4,\hspace{1ex}
hep-ph/0307108}

\title{\large\bf Wrong-sign Kaons in B decays and New Physics}

\author{Eung Jin Chun$^\dagger$ and  Jae Sik Lee$^*$}

\affiliation{$^\dagger$Korea Institute for Advanced Study\\
     207-43 Cheongryangri-dong, Dongdaemun-gu \\ Seoul 130-722, Korea\\
$^*$Department of Physics and Astronomy, University of Manchester,
Manchester M13 9PL, United Kingdom}


\begin{abstract}
New physics can possibly emerge in the $B$ decays into wrong-sign kaons
for which the standard model contributions are extremely suppressed.
We analyze two-body decays of $\bar{B}^0$ and $B^-$ mesons involving
the $b\to dd\bar{s}$  ($\Delta S=-1$) and $b\to ss\bar{d}$
($\Delta S=+2$) transitions in a model independent way, and 
examine various wrong-sign kaon signals  which are 
expected to be observed in the future $B$ experiments.  
Our analysis shows that 
it would be possible to identify the origin of
new physics  through the combined analysis of several 
$B$ decay modes involving one or two  wrong-sign $K^*$'s.
\end{abstract}



\maketitle

\end{titlepage}


\section{Introduction}

The standard processes of the $b$-quark decay to $s$-quark involve
the $\Delta S=0$ transition  $b \to d s\bar{s}$ coming
from Penguin diagrams, and
the $\Delta S=+1$ transitions $b \to s q\bar{q^\prime}$ ($q,q^\prime =u,c$)
induced from the tree-level $W^\pm$ exchange and $b \to s q\bar{q}$
($q=u,c,d,s$) from Penguin diagrams. Regarding these processes as
the ``right-sign'' $s$-quark (or kaon) decays of $B$ mesons, the
so-called ``wrong-sign'' decays can appear through the
$\Delta S=-1$ and $\Delta S=+2$ transitions,
\begin{equation}
b \to dd\bar{s} \quad(\Delta S=-1)\quad\mbox{or}\quad b \to ss\bar{d}
\quad(\Delta S=+2)\,.
\end{equation}
In the standard model (SM), such processes come from box
diagrams exchanging $W^\pm$ bosons in the loop inducing the effective
Lagrangian:
\begin{equation}
 {\cal L}_{eff}= {4 G_F\over\sqrt{2}} \left[
   C^{dd}_{SM}\, (\bar{d}_L\gamma_\mu b_L) (\bar{d}_L \gamma^\mu s_L)
  +C^{ss}_{SM}\, (\bar{s}_L\gamma_\mu b_L) (\bar{s}_L \gamma^\mu d_L)  \right]
\end{equation}
where the SM coefficients are exceedingly small due to the strong
GIM-suppression and the small CKM angles involved \cite{hlsz}.
Rough estimation gives $C^{dd}_{SM}\sim \lambda^8
G_Fm_W^2/2\sqrt{2}\pi^2$ and $C^{ss}_{SM}\sim \lambda^7 G_Fm_W^2/2\sqrt{2}\pi^2$
which make the inclusive branching ratios below $10^{-13}$ and
$10^{-11}$, respectively.  Such effects will be  beyond the reach
of any possible future experiments such as super-B
factory \cite{superB}, Tevatron \cite{tevb} or LHC \cite{LHCb}
which will produce about $10^{10}-10^{12}$ $B \bar{B}$ mesons.

Given the suppressed SM contribution, certain new physics beyond
the SM could give sizable contributions to the wrong-sign
$s$-quark transitions and thus alter various observables of the
$B$ decays predicted in the context of the SM.   Typical
examples of new physics such as two Higgs-doublet models and
supersymmetric standard model with squark flavor mixing or
R-parity  violation have been considered in Ref.~\cite{hlsz}. In
this paper, we investigate the effects of wrong-sign $s$-quark
operators on various physical observables in the $B$
decays, and  examine how to extract such effects in the future
experiments, without resorting to specific models of new physics.

\medskip

We start with introducing the most
general scalar and vector current
effective
Lagrangian for the $\Delta S=-1$ transition;
\begin{eqnarray} \label{wrong-op}
 {\cal L}_{eff}&=& {4 G_F \over \sqrt{2} } \left[
   S^{dd}_{LL}\, (\bar{d}_R b_L) (\bar{d}_R  s_L)
  +S^{dd}_{RR}\, (\bar{d}_L b_R) (\bar{d}_L  s_R)
\right.
\\
&&~~~~
 +S^{\prime dd}_{LL}\, (\bar{d}_R^\alpha b_L^\beta)
                       (\bar{d}_R^\beta  s_L^\alpha)
 +S^{\prime dd}_{RR}\, (\bar{d}_L^\alpha b_R^\beta)
                       (\bar{d}_L^\beta  s_R^\alpha)
\nonumber\\
&&~~~~
  +C^{dd}_{LL}\, (\bar{d}_L\gamma_\mu b_L) (\bar{d}_L\gamma^\mu s_L)
  +C^{dd}_{RR}\, (\bar{d}_R\gamma_\mu b_R) (\bar{d}_R \gamma^\mu s_R)
\nonumber\\
&&~~~~
  +C^{dd}_{LR}\, (\bar{d}_L\gamma_\mu b_L) (\bar{d}_R \gamma^\mu s_R)
  +C^{dd}_{RL}\, (\bar{d}_R\gamma_\mu b_R) (\bar{d}_L \gamma^\mu s_L)
\nonumber\\
&&~~~~ \left.
  +C^{\prime dd}_{LR}\, (\bar{d}_L^\alpha \gamma_\mu b_L^\beta)
                        ( \bar{d}_R^\beta \gamma^\mu s_R^\alpha)
  +C^{\prime dd}_{RL}\, (\bar{d}_R^\alpha \gamma_\mu b_R^\beta)
                        (\bar{d}_L^\beta \gamma^\mu s_L^\alpha)  \right]
  + h.c.\,,
\nonumber
\end{eqnarray}
where $\alpha$ and $\beta$ are color indices. Note that we omitted
the scalar operators of the $LR$ and $RL$ types,
the vector operators of the $LL$ and $RR$ types  and tensor
operators as they  can be rewritten in terms of the above scalar
and vector operators after Fierz transformations.
For the $\Delta S=-2$ transition,
one takes the exchange, $d\leftrightarrow s$.
Among the typical examples of new physics,
the largest possible coefficients may be obtained with R-parity violation
which gives rise to $C'_{RL}$ and $C'_{LR}$ at the tree-level as follows:
\begin{eqnarray} \label{Crpv}
&&
C^{\prime dd}_{LR} =
       \sum_n {\sqrt{2}\over8}{\lambda'_{n31}\lambda^{\prime *}_{n12}
               \over G_F m^2_{\tilde{\nu}_n} }\,,\quad
C^{\prime dd}_{RL} =
       \sum_n {\sqrt{2}\over8}{\lambda'_{n21}\lambda^{\prime *}_{n13}
               \over G_F m^2_{\tilde{\nu}_n} }\,,
\\
&&
C^{\prime ss}_{LR} =
       \sum_n {\sqrt{2}\over8}{\lambda'_{n32}\lambda^{\prime *}_{n21}
               \over G_F m^2_{\tilde{\nu}_n} }\,,\quad
C^{\prime ss}_{RL} =
       \sum_n {\sqrt{2}\over8}{\lambda'_{n12}\lambda^{\prime *}_{n23}
               \over G_F m^2_{\tilde{\nu}_n} }\,,
\nonumber
\end{eqnarray}
where $ m^2_{\tilde{\nu}_n}$ is the mass of the mediating
sneutrino of the $n$--th generation.  We define that the
R-parity violating couplings are given in the superpotential as
follows:
$$
W_{\lambda^\prime}=\lambda^\prime_{ijk}
(E_iV^\dagger_{jl}U_lD_k^c-L_i^0D_jD_k^c) \,,
$$
where $L_i=(L_i^0, E_i)$ and $Q_i=(U_i, D_i)$ are the lepton and
quark $SU(2)$ doublets, and $D_i^c$ is the $SU(2)$ singlet
anti-quark superfields.  Here $V_{ij}$ is the CKM matrix of quark fields.
Let us note that the $C^\prime_{LR,RL}$ couplings 
induced by R-parity violation
can be as large as $0.1-0.01$  within the present experimental bounds
\cite{hlsz,bhatt}.
In the following, we will take the above new physics coefficients,
$C^\prime_{LR,RL}$, for  specific illustrations.

\section{$\Delta S=-1$ Transition : $b \to dd\bar{s}$ }

\newcommand{\Bb}{\bar{B}}
\newcommand{\Kb}{\bar{K}}
\newcommand{\Ab}{\bar{A}}
\newcommand{\wb}{\bar{w}}

Let us first consider the two--body decays of the neutral and charged $B$
mesons into $\pi$ and $K$ mesons arising
from $\Delta S=-1$ transition:
$$
\Bb^0\to \pi^0 K^0\,, \quad \pi^0 K^{*0}
\quad\mbox{and}\quad
B^-\to \pi^- K^0\,, \quad \pi^- K^{*0} \,.
$$
Within the factorization framework \cite{ali},
we obtain the following amplitudes for the neutral $B$ meson decays:
\begin{eqnarray}
&&A(\Bb^0\to \pi^0 K^0) =  i { G_F \over{2}} f_{K}
(m_B^2-m_\pi^2) F_0^{B\to \pi}(m_{K^0}^2) \times
\\
&&~~~~~~~~~~~
\Bigg\{ \frac{1}{2}r_1
[(1-{1\over2}\xi)(S^{dd}_{LL}-S^{dd}_{RR})+
 (\xi -{1\over2})(S^{\prime dd}_{LL}-S^{\prime dd}_{RR}) ]
\nonumber\\
&&~~~~~~~~
       +[(1+\xi) (C^{dd}_{LL}-C^{dd}_{RR})
       +(r_1 \xi-1 )(C^{dd}_{LR}-C^{dd}_{RL})
 +(r_1-\xi) (C^{\prime dd}_{LR}-C^{\prime dd}_{RL}) ]
 \Bigg\} \,,
\nonumber\\
\vphantom{frac{\frac{A}{A}}{A}}
&&A(B^-\to \pi^- K^{0}) = \sqrt{2} A(\Bb^0\to \pi^0 K^{0}) \,,
\label{Q2}
\\
\vphantom{frac{\frac{A}{A}}{A}}
&& A(\Bb^0\to \pi^0 K^{*0}) = -{G_F} (\epsilon^*
\cdot p_\pi)  m_{K^*}\times
\\
&&~~~~~~~~
 \Bigg\{ {m_{K^*}\over m_B+m_\pi} f^T_{K^*} F_2^{B\to \pi}(m^2_{K^*})
\frac{1}{2}
 [\xi (S^{dd}_{LL}+S^{dd}_{RR}) +
  (S^{\prime dd}_{LL}+S^{\prime dd}_{RR})]
\nonumber\\
&&~~~~~~~~~
 - f_{K^*} F_1^{B\to \pi}(m^2_{K^*})
 [(1+\xi) (C^{dd}_{LL}+C^{dd}_{RR})
       + (C^{dd}_{LR}+C^{dd}_{RL})
       +\xi (C^{\prime dd}_{LR}+C^{\prime dd}_{RL})  ] \Bigg\} \,,
\nonumber\\
\vphantom{frac{\frac{A}{A}}{A}}
&&A(B^-\to \pi^- K^{*0}) = \sqrt{2} A(\Bb^0\to \pi^0 K^{*0}) \,,
\end{eqnarray}
where $\xi\equiv 1/N_c$,
$r_1\equiv 2 m_{K^0}^2 /(m_b-m_d)(m_s+m_d)$
and $\epsilon^*$ is the polarization vector of the vector meson $K^*$.
The definitions of various
form factors and their numerical values taken for our calculations
are summarized in the Appendix.

\medskip

For comparisons with the SM right-sign amplitudes,
we define, for each decay mode, the ratio
$w_{\pi K}$ of the  wrong-sign (WS) amplitude with $\Delta S=-1$
to the corresponding SM one with  $\Delta S=+1$ 
(driven by the $b\to sd\bar{d}$ transition) as:
\begin{equation}
w_{\pi K} \equiv
\frac{A_{WS}(\Delta S=-1)}{A_{\rm SM}(\Delta S=+1)}\,.
\label{wbij1}
\end{equation}
In Table~I, we show the values of  $w_{\pi K}$ for each $\Delta S=-1$
operator defined in Eq.~(\ref{wrong-op}), taking $N_c=3$ and
$S^{(\prime)dd}_{II}, C^{(\prime)dd}_{IJ}=
|V_{tb}V_{ts}^*(a_4-a_{10}/2)|\simeq 1.54\times10^{-3}$.
For the numerical values used and definitions of
the coefficients $a_i$, {\it etc}, see Ref.~\cite{JKK,ali}.
Table~I clearly shows that there can be significant effects if
$S^{(\prime)dd}_{II} \sim 10^{-3}$ or
$C^{(\prime)dd}_{IJ} \sim 10^{-3}$ is allowed.

\medskip

As pointed out in Ref.~\cite{hlsz}, the $\Bb \to \pi K^{*0}$
mode will play major role for probing or constraining the
$\Delta S=-1$ transition.  Here, $\Bb$ denotes $\Bb^0$ or $B^-$ and
correspondingly $\pi$ can be $\pi^0$ or $\pi^-$.
Let us note that the  ratio $w_{\pi K^{*0}} = 
 A_{WS}(\Bb\to \pi K^{*0})/ A_{SM}(\Bb\to \pi \Kb^{*0})$
can be determined by comparing two
branching fractions of $\Delta S = -1$ and $\Delta S = +1$
transitions:
\begin{equation} \label{wpiK*}
 |w_{\pi K^{*0}}|^2 = { {\cal B}(\Bb \to \pi K^{*0}\to \pi\,\pi^- K^+) \over
               {\cal B}(\Bb \to \pi \Kb^{*0}\to \pi\, \pi^+ K^-)  } \,.
\end{equation}
The current measurements at the $B$ factories have started to constrain
$w_{\pi^- K^{*0}}$ which takes particularly simple form as
\begin{equation}
 w_{\pi^- K^{*0}} = \Bigg\{ (1+\xi) [C^{dd}_{LL}+C^{dd}_{RR}]
  +  [C^{dd}_{LR}+C^{dd}_{RL}]
  +  \xi [C^{\prime dd}_{LR}+C^{\prime dd}_{RL}] \Bigg\} {1\over C_{SM}}
\end{equation}
where $\xi=1/N_c$ and $C_{SM}=  V_{tb}V^*_{ts}(a_4-a_{10}/2)$.
Here we neglected the contributions from the scalar operators.
Recently, accumulating about $(6-9)\times 10^7$ $B\bar{B}$ pairs,
BaBar Collaboration reported the measurements;
\begin{eqnarray}
{\cal B}(B^-\to \pi^-\pi^+ K^{-}) &=& (59.1\pm 3.8 \pm 3.2)\times10^{-6}
 \quad   \mbox{\cite{baba1}} \,,
\nonumber\\
{\cal B}(B^- \to \pi^- \Kb^{*0}\to \pi^- \pi^+ K^-) &=&
(10.3\pm 1.2^{+1.0}_{-2.7}) \times10^{-6} \quad~~~ \mbox{\cite{baba2}} \,.
\end{eqnarray}
With this, the measured upper bound  of the
3-body wrong-sign branching ratio,
\begin{equation}
{\cal B}(B^-\to \pi^-\pi^- K^{+}) < 
1.8\times10^{-6} \quad \mbox{\cite{baba1}} \,,
\end{equation}
is translated to the bound,
\begin{equation}
{\cal B}(B^-\to \pi^-K^{*0} \to \pi^- \pi^- K^+) 
< 3.1\times 10^{-7} \,.
\end{equation}
%
Therefore, from Eq.~(\ref{wpiK*}) one obtains the bound,
\begin{equation} \label{bound1}
|w_{\pi^- K^{*0}}|< 0.17 \,.
\end{equation}
This implies that the coefficient $C^\prime_{LR,RL}$ gets the
constraint of $|C^\prime| \lesssim 0.17 N_c |C_{SM}| =
7.9\times10^{-4}(N_c/3)$.
Applying this to the R-parity violation in Eq.~(\ref{Crpv}),  
we obtain the following stringent bound:
\begin{equation}
|\lambda'_{n31}\lambda^{\prime *}_{n12}|\,,\;
|\lambda'_{n21}\lambda^{\prime *}_{n13}| < 5.2\times10^{-4} 
\left( m_{\tilde{\nu}_n} \over 100 \mbox{ GeV} \right)^2  .
\end{equation}
Considering future experiments producing $10^{11}$
$B$ mesons,  it is expected to probe $|w_{\pi^- K^{*0}}|$ below
the level of 1 \%, providing the limit on the coefficients
$C$'s  down to $3\times10^{-5}$.

\medskip

The decay into two pseudoscalar mesons is also useful to probe
$\Delta S=-1$ though it is more difficult compared with 
the $\pi K^{*0}$ mode.
First of all, the presence of the wrong--sign operators affects the experimental
determination of the branching ratio of the mode $\bar{B} \to \pi \bar{K}^0$
from the measurements of ${\cal B}(\bar{B} \to \pi K^0_{S,L})$.
Both the $\Delta S=-1$ transition, $\bar{B} \to \pi K^0$, and
the $\Delta S=+1$ one,  $\bar{B} \to \pi \bar{K}^0$,
contribute to the decays $\bar{B} \to \pi K^0_{S,L}$ through the
$K$--$\bar{K}$ mixing.  
The  amplitude of the $B$ decay into $\pi K_S$ or $\pi K_L$
is  given by
\begin{equation} \label{piKsl}
 \Ab_{\pi K^0_{S,L}} = p_K \Ab_{\pi K^0} \pm q_K \Ab_{\pi \bar{K}^0}\,,
\end{equation}
where $\bar{A}_{M_1 M_2}\equiv A(\Bb\to M_1 M_2)$.
In Eq.~(\ref{piKsl}),
$p_K$ and $q_K$ are the coefficients relating
the $K$ meson mass eigenstates with the flavor eigenstates;
\begin{equation}
 |K_{S,L}\rangle  = p_K |K^0\rangle \pm q_K |\bar{K}^0\rangle \,.
\end{equation}
Recall that $p_k,q_K= (1\pm \bar{\epsilon})/\sqrt{2(1+|\bar{\epsilon}|^2)}$
with $|\bar{\epsilon}| \sim 10^{-3}$.
Denoting  the ratio of the wrong-sign
amplitude to the SM one as
$  w_{\pi K^0} = \Ab_{\pi K^0}/ \Ab_{\pi \Kb^0} $,
we get
\begin{equation} \label{piK}
 2 {\cal B}(\Bb \to \pi K^0_{S,L}) =
 {\cal B}(\Bb \to \pi \Kb^0)_{\rm SM} 
 \Big|1 \pm {p_K\over q_K}w_{\pi K^0}\Big|^2 \,,
\end{equation}
where $p_K/q_K \simeq 1$.
The SM relation, ${\cal B}(\Bb \to \pi \Kb^0)=
2{\cal B}(\Bb \to \pi K^0_{S,L})$, can obviously be
invalidated in the presence of
the wrong-sign amplitudes, and thus it has to be checked experimentally.
Current experiments at $B$ factories only look for the modes $\Bb
\to \pi K^0_S$.  The present world average of the $\Bb\to \pi K_S$
 branching ratios are \cite{tomura}:
\begin{eqnarray} \label{data1}
2\,{\cal B}(\Bb^0\rightarrow \pi^0 K^0_S )&=&
  (11.5 \pm 1.7)\times 10^{-6}\,,
\nonumber \\
2\,{\cal B}(B^-\rightarrow \pi^- K^0_S )&=&
 (20.6 \pm 1.4)\times 10^{-6}\,.
\end{eqnarray}
This  can be compared with the  SM prediction:
$2\,{\cal B}(\Bb^0\rightarrow \pi^0 K^0_S )_{SM} =5.1\times10^{-6}$ and
$2\,{\cal B}(B^-\rightarrow \pi^- K^0_S )_{SM} =15\times10^{-6}$,
which are derived within the factorization scheme taking
the standard values for the input parameters as specified in Appendix
and $\xi=1/3$.  The apparent discrepancies between the experimental
and theoretical values can be cured by the wrong-sign amplitude contribution
as in Eq.~(\ref{piKsl}).
With the results of Table 1, the bound (\ref{bound1}) can be translated to
$|w_{\pi^0 K^0}| < 0.13$,  and $|w_{\pi^- K^0}| < 0.11$ for the case of
the new physics  coupling $C^\prime_{LR,RL}$.  Thus, the maximal
contributions of the new physics (NP) to $w_{\pi K}$ can give a better
explanation of the data as we get
$2\,{\cal B}(\Bb^0\rightarrow \pi^0 K^0_S )_{NP} =6.5\times10^{-6}$ and
$2\,{\cal B}(B^-\rightarrow \pi^- K^0_S )_{NP} =18\times10^{-6}$.
However, it is premature to make any definite conclusion about the role
of the wrong-sign amplitudes
since the theoretical calculations  have large uncertainties not only within
the factorization scheme \cite{ali} but also in
any other approaches \cite{chen,bbns,du,keum}.

In relation to this, let us
remark on  the ``wrong-sign'' kaon contribution  to
the isospin violation  \cite{trojan} in the  $B\to \pi K$ modes;
$$ \Bb^0 \to \pi^0 K_S\,,\; \pi^- K^+ \,;\quad
   B^- \to \pi^- K_S\,,\; \pi^0 K^- .  $$
As discussed, the $\Delta S=-1$ operators contribute only
to $  \Bb^0 \to \pi^0 K_S$ and  $B^- \to \pi^- K_S$ as in Eq.~(\ref{Q2}).
This shows that the experimental data (\ref{data1}), implying
$2{\cal B}(\Bb^0\rightarrow \pi^0 K^0_S )>
{\cal B}(B^-\rightarrow \pi^- K^0_S )$, can be explained by an
enhanced electro-weak penguin contribution coming from new physics
as analyzed in a recent paper \cite{yoshi}.  In fact, all the isospin
violating relations could be a consequence of both the electro-weak
penguin and the wrong-sign amplitude \cite{alex}.

\medskip

A direct way to probe the above wrong-sign amplitude, $w_{\pi K^0}$,
is to reconstruct $K_L$ experimentally.
This allows us to measure 
the following rate asymmetries \cite{bigi} which are nearly
vanishing in the SM:
\begin{equation} \label{Asl}
  \bar{\cal A}^{\pi}_{SL} \equiv
  { \Gamma(\Bb \to \pi K^0_{S}) -\Gamma(\Bb \to \pi K^0_{L})
   \over \Gamma(\Bb \to \pi K^0_{S}) + \Gamma(\Bb \to \pi K^0_{L})  }
           = { 2 {\rm Re}(w_{\pi K^0}) \over 1+ |w_{\pi K^0}|^2 }\,.
\end{equation}
In the current $B$ factories, only the direction of $K^0_L$
can be measured.  Then, its
momentum can be calculated from the $B$--mass constraint to
reconstruct the mode $\Bb \to \pi K^0_L$. This is the way to
measure CP violation in the $\Bb \to J/\psi K^0_L$ mode.
Contrary to the $J/\psi K_L^0$ case,  the final state $\pi K^0_L$
suffers from a huge background which makes it hard  to separate
out the candidate events.  In order to avoid it, one may have to
fully reconstruct the other $B$, by which, however, we can only
use about 0.1\% of the produced $B \Bb$ pairs. Considering the
branching ratio $\approx 10^{-5}$ of the $\pi K^0$ mode and
the 0.1\% detection efficiency, one can collect about
1000 $\pi K^0_L$ events from $10^{11}$ $B \bar{B}$ pairs.
Therefore, the $K_S$--$K_L$ asymmetry at the level of
a few \% could be seen in  the future experiments.

\medskip

The presence of the  wrong-sign amplitude can appear also in
the direct CP asymmetry of $B^\pm \to \pi^\pm K^0_S$. In
the SM, the CP asymmetry
\begin{equation}
 {\cal A}_{CP} = { \Gamma(B^+ \to \pi^+ K^0_S) - \Gamma(B^- \to \pi^- K^0_S)
   \over \Gamma(B^+ \to \pi^+ K^0_S) + \Gamma(B^- \to \pi^- K^0_S) }
\end{equation}
is expected to be of order 1\% arising from the interference of
two penguin contributions (to $\bar{A}_{\pi \Kb^0}$ in Eq.~(\ref{piKsl}))
with the CKM-suppressed relative amplitude $\sim 0.02$.  The
wrong-sign amplitudes, $A_{\pi^+ \Kb^0}$ for $B^+$ and $\Ab_{\pi^-
K^0}$ for $B^-$ as in Eq.~(\ref{piKsl}), can give rise to another
interfering effect. Under the condition that the wrong-sign
amplitudes dominates over the CKM-suppressed penguin amplitudes,
we obtain
\begin{equation} \label{Acp}
 {\cal A}_{CP}  = { 2 |w_{\pi^- K^0}| \sin\Delta\phi \sin\Delta\delta
     \over 1 + 2 |w_{\pi^- K^0}| \cos\Delta\phi \cos\Delta\delta +
      |w_{\pi^- K^0}|^2  }\,,
\end{equation}
where $\Delta\phi$ ($\Delta\delta$) is the relative weak (strong)
phase of the right and wrong sign amplitudes.
At the moment, the above CP asymmetry is measured with the
accuracy of 10-20\% \cite{tomura}  which puts a
constraint on $|w_{\pi^- K^0}|$  (with $\Delta\phi, \Delta\delta \sim 1$ )
close  to the bound, $|w_{\pi^- K^0}|<0.11$,
coming from the branching ratio measurements discussed below
Eq.~(\ref{wpiK*}).  The CP asymmetry  ${\cal A}_{CP}$ is expected to be
improved to the level of one percent in the future experiments, and thus
could  provide an indirect way to probe the wrong-sign amplitudes.

\section{$\Delta S=+2$ Transition : $b \to ss\bar{d}$}

In this section, we discuss the two--body decays of the charged and
neutral $B$ mesons induced by the $\Delta S=+2$ operators:
\begin{eqnarray}
{\rm PP~~modes}\,\,&:&\quad \Bb^0\to \Kb^0 \Kb^0\,, \quad B^-\to K^- \Kb^0\,,
\nonumber \\
{\rm PV~~modes}\,\,&:&\quad \Bb^0\to \Kb^0 \Kb^{*0}\,,
\quad B^-\to K^- \Kb^{*0}\,, \quad B^-\to K^{*-} \Kb^0\,,
\nonumber\\
{\rm VV~~modes}\,\,&:&\quad \Bb^0\to \Kb^{*0} \Kb^{*0}\,,
\quad B^-\to K^{*-} \Kb^{*0}\,. \nonumber
\end{eqnarray}
The amplitudes for the decay modes into two pseudoscalar mesons
(PP), a pseudoscalar and a vector mesons (PV), and two vector
mesons (VV) are given as follows:
\begin{eqnarray}
&&A(\Bb^0 \to \Kb^0 \Kb^0) =
 i \sqrt{2} G_F f_{K} (m_B^2-m_{K^0}^2) F_0^{B\to K}(m_{K^0}^2)\times
\\
&&\quad
\Bigg\{ {r_2\over2} [(1-{1\over2}\xi)(S^{ss}_{LL}-S^{ss}_{RR}) +
(\xi-{1\over2})(S^{\prime  ss}_{LL}-S^{\prime ss}_{RR})]
\nonumber\\
&&\quad + [ (1+\xi)(C^{ss}_{LL}-C^{ss}_{RR})
 +(r_2\xi-1 )(C^{ss}_{LR}-C^{ss}_{RL})
 +(r_2-\xi) (C^{\prime ss}_{LR}-C^{\prime ss}_{RL})]\Bigg\}\,,
\nonumber\\
\vphantom{frac{\frac{A}{A}}{A}}
&& A(B^-\to K^- \Kb^{0}) = {1\over2} A(\Bb^0\to \Kb^0 \Kb^0)\,,
\\
\vphantom{frac{\frac{A}{A}}{A}} 
&& A(\Bb^0\to \Kb^0 \Kb^{*0})  =
 -\sqrt{2} G_F (\epsilon^*\cdot p_B) m_{K^*}\times
 \\
&&\quad
 \Bigg\{ {m_{K^*}\over m_B+m_K} f^T_{K^*} F_2^{B\to
K}(m^2_{K^*}) {1\over2}[\xi (S^{ss}_{LL}+S^{ss}_{RR}) + (
S^{\prime ss}_{LL}+S^{\prime ss}_{RR})]
\nonumber\\
&&\quad
 + {r_3\over2} f_K A_0^{B\to K^*}(m^2_{K^0})
[(1-{1\over2}\xi)(S^{ss}_{LL}+S^{ss}_{RR})+
(\xi-{1\over2})(S^{\prime ss}_{LL}+S^{\prime ss}_{RR})]
\nonumber\\
&&\quad
- f_{K^*} F_1^{B\to K}(m^2_{K^*}) [
(1+\xi)(C^{ss}_{LL}+C^{ss}_{RR})
 + (C^{ss}_{LR}+C^{ss}_{RL})
 + \xi (C^{\prime ss}_{LR}+C^{\prime ss}_{RL}) ]
\nonumber\\
&&\quad
  - f_{K} A_0^{B\to K^*}(m^2_{K^0}) [
(1+\xi)(C^{ss}_{LL}+C^{ss}_{RR})
\nonumber\\
&& \qquad\qquad\qquad\qquad\quad
  -(1+r_3 \xi)(C^{ss}_{LR}+C^{ss}_{RL})
        -(\xi+r_3) (C^{\prime ss}_{LR}+C^{\prime ss}_{RL}) ]
        \Bigg\}\,,
\nonumber\\
\vphantom{frac{\frac{A}{A}}{A}} 
&& A(B^-\to K^- \Kb^{*0})  =
 - \sqrt{2} G_F (\epsilon^*\cdot p_B) m_{K^*} \times
 \\
&&\quad
 \Bigg\{ {m_{K^*}\over m_B + m_K} f^T_{K^*} F_2^{B\to
K}(m^2_{K^*})  {1\over2} [ \xi(S^{ss}_{LL}+S^{ss}_{RR}) +
 (S^{\prime ss}_{LL}+S_{RR}^{\prime ss})]
\nonumber\\
&&\quad
 - f_{K^*} F_1^{B\to K}(m^2_{K^*})   [
(1+\xi)(C^{ss}_{LL}+C^{ss}_{RR})
  + (C^{ss}_{LR}+C^{ss}_{RL})
        +\xi (C^{\prime ss}_{LR}+C^{\prime ss}_{RL}) ] \Bigg\}\,,
\nonumber\\
\vphantom{frac{\frac{A}{A}}{A}} && A(B^-\to K^{*-} \Kb^{0}) =
 - \sqrt{2} G_F (\epsilon^*\cdot p_{B}) m_{K^*}\,,
 f_K A_0^{B\to K^*}(m^2_{K^0})\times
\\
&&\quad
 \Bigg\{ {r_3\over2}[(1-{1\over2}\xi)(S^{ss}_{LL}+S^{ss}_{RR}) +
(\xi-{1\over2})(S^{\prime ss}_{LL}+S^{\prime ss}_{RR}) ]
\nonumber\\
 &&\quad
    - [ (1+\xi)(C^{ss}_{LL}+C^{ss}_{RR})
  -(1+r_3 \xi)(C^{ss}_{LR}+C^{ss}_{RL})
        -(\xi+r_3) (C^{\prime ss}_{LR}+C^{\prime ss}_{RL}) ]
        \Bigg\}\,,
\nonumber\\
\vphantom{frac{\frac{A}{A}}{A}}
&& A(\Bb^0\to \Kb^{*0} \Kb^{*0}) =
\\
&&\quad
- \sqrt{2}G_F f^T_{K^*} \Bigg\{ \Big[(\epsilon_{\mu\nu\rho\sigma}
 \epsilon_1^{*\mu} \epsilon_2^{*\nu} p_1^\rho p_2^\sigma) T_1^{B\to
 K^*}(m^2_{K^*})\Big] \Big[ \xi(S^{ss}_{LL}+S^{ss}_{RR}) +
 (S^{\prime ss}_{LL}+S^{\prime ss}_{RR}) \Big]
 \nonumber\\
&&\quad
~~  - i \Big[ (\epsilon^*_1\cdot p_2)(\epsilon^*_2\cdot p_1)
 \Big(T_2^{B\to K^*}(m^2_{K^*}) + {m^2_{K^*} \over m^2_B-m^2_{K^*} }
  T_3^{B\to K^*}(m^2_{K^*}) \Big) -
\nonumber\\
&& \quad
~~~~  {(\epsilon^*_1\cdot \epsilon^*_2)\over2}(m^2_B-m^2_{K^*})
 T_2^{B\to K^*}(m^2_{K^*})  \Big]
\Big[ \xi(S^{ss}_{LL}-S^{ss}_{RR})
 + (S^{\prime ss}_{LL}-S^{\prime ss}_{RR}) \Big]
\Bigg\}
\nonumber\\
&& \quad + \sqrt{2} G_F m_{K^*} f_{K^*} \Bigg\{
 \Big[(\epsilon_{\mu\nu\rho\sigma}
 \epsilon_1^{*\mu} \epsilon_2^{*\nu} p_1^\rho p_2^\sigma)
 {2 V^{B\to K^*}(m^2_{K^*}) \over m_B + m_{K^*} } \Big] \times
\nonumber\\
&&\quad\qquad
\Big[(1+\xi)(C^{ss}_{LL}+C^{ss}_{RR})+(C^{ss}_{LR}+C^{ss}_{RL})
 +\xi(C^{\prime ss}_{LR}+C^{\prime ss}_{RL}) \Big]
\nonumber\\
&&\quad\qquad
 -i \Big[(\epsilon^*_1 \cdot \epsilon^*_2)
 (m_B + m_{K^*}) A_1^{B\to K^*}(m^2_{K^*})
  - (\epsilon^*_1\cdot p_2)(\epsilon^*_2\cdot p_1)
\frac{2A_2^{B\to K^*}(m^2_{K^*})}{m_B+m_{K^*}} \Big]
\times
 \nonumber\\
&&\quad\qquad
 \Big[(1+\xi)(C^{ss}_{LL}-C^{ss}_{RR})+(C^{ss}_{LR}-C^{ss}_{RL})
  +\xi(C^{\prime ss}_{LR}-C^{\prime ss}_{RL})\Big] \Bigg\}\,,
\nonumber\\
\vphantom{frac{\frac{A}{A}}{A}}
&& A(B^-\to K^{*-} \Kb^{*0})  = {1\over2} A(\Bb^0\to \Kb^{*0}
\Kb^{*0})\,,
\end{eqnarray}
where $r_2\equiv 2m_{K^0}^2 /(m_b-m_s)(m_s+m_d)$, $r_3\equiv
2m_{K^0}^2 /(m_b+m_s)(m_s+m_d)$ and the index 1 or 2 labels each
$\bar{K}^{*0}$ in the $\Bb^0 \to \Kb^{*0} \Kb^{*0}$ mode.

\newcommand{\wbh}{\widehat{\bar{w}}}

Again, Table I shows the ratio
$w_{KK} $ of the wrong-sign $\Delta S=+2$ amplitudes to the corresponding
$\Delta S=0$ SM ones (driven by the $b\to ds\bar{s}$
transition) as:
\begin{equation}
w_{KK}\equiv
\frac{A_{WS}(\Delta S=+2)}{A_{\rm SM}(\Delta S=0)}\,,
\label{wbij2}
\end{equation}
contributed by each  $\Delta S=+2$ operator with the coefficient
$C^{(\prime)ss}_{IJ}$ or $S^{(\prime)ss}_{II}$ where $I,J =L,R$.
The numerical values are taken with the choice of
$N_c=3$ and $S^{(\prime)ss}_{II}, C^{(\prime)ss}_{IJ}
=|V_{tb}V_{td}^*(a_4-a_{10}/2)|=2.91\times10^{-4}$.
For the VV modes, we show the square-rooted ratio,  $|w_{K^* K^*}| \equiv
\sqrt{\Gamma_{WS}(\Delta S=2)/\Gamma_{SM}(\Delta S=0)}$
since the direct comparison between amplitudes is not possible.
In the following, we closely examine phenomenological implications
of the wrong--sign $\Delta S=+2$ transition in each mode.

\smallskip

$\bullet$ {\bf PP modes} :

\noindent
When there exist the wrong-sign amplitude
of the process $\Bb^0 \to \Kb^0 \Kb^0$ as well as 
the SM amplitude  of the process $\Bb^0 \to K^0 \bar{K}^0$,
the final states $|K_A;K_B\rangle$ with $A,B=S,L$ can be written as
\begin{eqnarray}
 |K_{ S,L }; K_{S,L}\rangle
 &=& + q_K^2 |\Kb^0;\Kb^0\rangle
 \pm p_K q_K \left( |K^0;\bar{K}^0\rangle
    + |\bar{K}^0;K^0\rangle  \right)\,,
\nonumber\\
 |K_{S,L}; K_{L,S}\rangle
 &=& -q_K^2 |\Kb^0; \Kb^0\rangle \mp p_K q_K
     \left( |K^0; \bar{K}^0\rangle
    - |\bar{K}^0; K^0\rangle  \right)\,,
\end{eqnarray}
where the first and second $K$'s are to be labeled by its momentum
$\vec{k}$ and $-\vec{k}$, respectively, in the $\Bb^0$ rest frame.
Note that we have neglected the state $|K^0;K^0\rangle$ as it is
irrelevant for our discussion. Rotation invariance implies that
the antisymmetric combination of two different $K$'s vanishes in
the amplitude.
Thus, we obtain the following amplitudes with
symmetrized final states:
\begin{eqnarray}
 \Ab_{K_{S}K_{S}} &=& q_K^2 \Ab_{\Kb^0 \Kb^0}
          + \sqrt{2} p_K q_K \Ab_{K^0\bar{K}^0}\,,
\nonumber\\
 \Ab_{K_{L}K_{L}} &=& q_K^2 \Ab_{\Kb^0 \Kb^0}
          - \sqrt{2} p_K q_K \Ab_{K^0\bar{K}^0}\,,
\\
 \Ab_{K_{S}K_{L}} &=&  -\sqrt{2} q_K^2  \Ab_{\Kb^0 \Kb^0}\,,
\nonumber
\end{eqnarray}
for the $\bar{B}^0$ decays to $K_SK_S$, $K_L K_L$ and $K_S K_L$,
respectively.
%
%
We see that the SM predictions, $\Gamma(K_SK_S) = \Gamma(K_L K_L)$
and $\Gamma(K_S K_L)=0$, are modified as
\begin{eqnarray}
 \Gamma(\Bb^0\to K_S K_S)&=&\Gamma(\Bb^0\to K_S K_S)_{\rm SM}
\,\Big| 1+{1\over\sqrt{2}} {q_K\over p_K} w_{\Kb^0 \Kb^0} \Big|^2\,,
\nonumber\\
 \Gamma(\Bb^0\to K_L K_L)&=&\Gamma(\Bb^0\to K_S K_S)_{\rm SM}
\,\Big| 1-{1\over\sqrt{2}} {q_K\over p_K} w_{\Kb^0 \Kb^0} \Big|^2\,,
\\
 \Gamma(\Bb^0\to K_S K_L)&=&\Gamma(\Bb^0\to K_S K_S)_{\rm SM}
\,\Big| {q_K\over p_K} w_{\Kb^0 \Kb^0} \Big|^2\,.
\nonumber
\end{eqnarray}
where
$w_{\Kb^0 \Kb^0}= \Ab_{\Kb^0 \Kb^0}/ \Ab_{K^0\bar{K}^0}$.

The best way to observe $\Delta S=+2$ transition is to measure
the following observables by reconstructing $K_L$ experimentally:
\begin{eqnarray}
\bar{\cal R}_{SL} &\equiv&
 { \Gamma(\Bb^0 \to K_S K_L) \over \Gamma(\Bb^0 \to K_S K_S) } =
 {|w_{\Kb^0\Kb^0}|^2 \over | 1+{1\over\sqrt{2}} w_{\Kb^0\Kb^0} |^2}\,,
\nonumber\\
\bar{\cal A}^K_{SL} &\equiv&
  { \Gamma(\Bb^0 \to K_S K_S) - \Gamma(\Bb^0 \to K_L K_L) \over
     \Gamma(\Bb^0 \to K_S K_S) + \Gamma(\Bb^0 \to K_L K_L)}
   = {2\sqrt{2} \mbox{Re}(w_{\Kb^0\Kb^0}) \over 2 + |w_{\Kb^0\Kb^0}|^2 }\,.
\end{eqnarray}
The observable $\bar{\cal R}_{SL}$ is of a particular interest as
it measures the absolute value of the wrong-sign amplitude.
Following the similar argument below Eq.~(\ref{Asl}), we find that
 $\bar{\cal R}_{SL}$ could be measured up to the level of
10 \% with $10^{11} B\bar{B}$ mesons taking the branching ratio
${\cal B}(\bar{B}^0\to K^0\bar{K}^0)_{SM} \sim 10^{-6}$.
 Note that the current experimental results give the bound;
$2{\cal B}(\bar{B}^0\to K_S K_S) < (1.6-3.2) \times 10^{-6}$
\cite{tomura}.
However, it will be almost impossible to measure $\bar{\cal
A}^K_{SL}$  involving two $K_L$ final state.  We expect that
precision measurements of the above observables can be made if
future $B$ experiments are equipped with a hadronic calorimetry
with a significant ability of reconstructing $K_L$ which is not
anticipated in the present plans.

\medskip

For the decays $B^- \to K^- K_{S,L}$,  we can get the similar expressions
as in Eqs.~(\ref{Asl}) and (\ref{Acp}) by replacing $\pi^-$
with $K^-$ and $(p_K/q_K)w_{\pi^- K^0}$ with $(q_K/p_K)w_{K^-\Kb^0}$.
Note that the CP asymmetry ${\cal A}_{CP}$  of $B^\pm \to K^\pm K_S$
is known to be about 20\% and thus the wrong-sign contribution has to be 
fairly large to see a new physics effect.

\smallskip

$\bullet$ {\bf  PV modes} :

\noindent
As discussed in the previous section, the production of $K^{*0}$ or
$\Kb^{*0}$ in the $B$ decays provides a straightforward way to
identify the right-sign or wrong-sign signals.
Let us first consider the $\Bb^0$ decays.  As shown in Table I,
the wrong-sign amplitude for  $\Bb^0\to \Kb^0\Kb^{*0}$ can be
compared with two right-sign amplitudes for  $\Bb^0\to \Kb^0 K^{*0}$
and $K^0 \Kb^{*0}$.  Here, it is amusing to note that the latter right-sign
amplitude exhibits a cancellation among SM contributions
thus one predicts
${\cal B}(K_S \Kb^{*0})_{SM}/{\cal B}(K_S K^{*0})_{SM} \lesssim 0.1$
\cite{ali,chen,bbns,du,keum}.
Now that the wrong-sign amplitude contributes only to the
$K_S \Kb^{*0}$ mode, it can alter the above SM prediction.  Namely,
the observation of
\begin{equation} \label{PV1}
 { {\cal B}(\Bb^0 \to K_S \Kb^{*0}) \over {\cal B}(\Bb^0 \to K_S K^{*0}) } >0.1
\end{equation}
will clearly be a signal for new physics inducing the wrong-sign amplitude.

In the case of $B^-$ decays, there are two ways of identifying
the wrong-sign amplitude.  One is to look for $B^- \to K^- \Kb^{*0}
\to K^-K^-\pi^+$ which is almost absent in the SM.
Another way is to observe $B^- \to K_S K^{*-}$. As in the $\Bb^0$ case,
the SM amplitude of $B^- \to K^0 K^{*-}$ is similarly suppressed
and thus the wrong-sign amplitude may have a larger contribution. As
a result,
\begin{equation} \label{PV2}
 { 2{\cal B}(B^- \to K_S K^{*-}) \over {\cal B}(B^- \to K^- K^{*0}) }>0.1
\end{equation}
may arise together with Eq.~(\ref{PV1}).

If the branching ratios of $B^-\to K^-\Kb^{*0},  K_S K^{*-}$ and
$\Bb^0\to K_S \Kb^{*0}$ are measured above the SM predictions,
it will be possible to identify which operators in Eq.~(\ref{wrong-op})
contribute to the wrong-sign amplitude.  To get an idea, let us compare
the branching ratios assuming one type of the coefficients, $C^{(\prime)}$
and $S^{(\prime)}$ exists:
\begin{eqnarray}
 {{\cal B}(B^-\to K^-\Kb^{*0}) \over {\cal B}(B^-\to K_S K^{*-})}
 &=& 3.8\, (C_{II}),\;\; 2.4\, (C_{IJ}), \;\; 0.35\, (C^\prime_{IJ}),\;\;
   0.024\, (S_{II}), \;\; 0.54\, (S^\prime_{II})\,,   \nonumber
\\
 {{\cal B}(\Bb^0\to K_S \Kb^{*0}) \over {\cal B}(B^-\to K_S K^{*-} )}
 &=& 5.7\, (C_{II}),\;\; 0.01\, (C_{IJ}), \;\; 0.34\, (C^\prime_{IJ}),\;\;
   1.2\, (S_{II}), \;\; 0.4\, (S^\prime_{II})\,,
\end{eqnarray}
where $I,J=L,R$ and $I \neq J$.
This shows, for instance,  that we should find the ratio
$ {\cal B}(K_S K^{*-}) : {\cal B}(K^-\Kb^{*0}) :
{\cal B}(K_S \Kb^{*0}) \simeq 3 : 1: 1$ if the R-parity violation
is the source of the wrong-sign amplitude as in Eq.~(\ref{Crpv}).

\smallskip

Considering again the $K_L$ measurement,
$\Gamma(K_{S}\bar{K}^{*0})\neq \Gamma(K_{L}\bar{K}^{*0})$
can arise due to the interference between the right and wrong sign
amplitudes, whose ratio $q_K \bar{A}_{\Kb^0 \Kb^{*0}}/p_K \bar{A}_{K^0
\Kb^{*0}}$ can be separated out by measuring the $K_S$--$K_L$
asymmetry in the decay  $\Bb^0 \to \Kb^{*0} K_{S,L}$. This has to
be contrasted with the modes, $\Bb^0 \to  K_{S,L}{K}^{*0}$, in
which no wrong-sign amplitude can interfere, and therefore, the SM
prediction $\Gamma(K_{S}{K}^{*0})= \Gamma(K_{L}{K}^{*0})$
persists.

\smallskip

$\bullet$ {\bf VV modes} :

\noindent
Having two $K^*$s in the final states, these modes also provide
a clean way to identify the wrong-sign signals \cite{hlsz}.
In the $K^*K^*$ modes, the standard right-sign  processes contain
two opposite-sign $K$'s: $\Bb^0 \to {K}^{0*} \bar{K}^{0*} \to
(K^+\pi^-)(K^-\pi^+)$ and  $B^- \to {K}^{*-} {K}^{*0} \to (K^-
\pi^0)(K^+ \pi^-)$.  On the other hand,  the wrong-sign processes
give rise to two same-sign $K$'s in the final states: $\Bb^0 \to
\bar{K}^{*0} \bar{K}^{*0} \to (K^-\pi^+)(K^-\pi^+)$ and  $B^- \to
{K}^{*-} \bar{K}^{*0} \to (K^- \pi^0)(K^- \pi^+)$.

\smallskip

It is worthwhile to look into the ratio of branching ratios of the decay modes
$\Bb^0\to \Kb^{*0}\Kb^{*0}$ (or  $B^-\to \Kb^{*-}\Kb^{*0}$)
and $B^-\to K^-\Kb^{*0}$,
which are almost forbidden in the SM framework. 
We observe that the contributions
to ${\cal B}(B^-\to K^-\Kb^{*0})$ from the scalar operators are suppressed by
the factor $\sim (m_{K^*}/2\,m_B)^2$
comparing with those from the vector operators.
On the other hand, the contributions to ${\cal B}(B^-\to \Kb^{*0}\Kb^{*0})$
from the scalar operators are not much different from those from the vector
operators. Specifically we find
\smallskip
\begin{eqnarray}
 {{\cal B}(\Bb^0\to \Kb^{*0}\Kb^{*0}) \over {\cal B}(B^-\to K^-\Kb^{*0})} =
340\,(S^{(\prime)}_{II})\,,\;\;
1.9\,(C^{(\prime)}_{IJ})\,,
\end{eqnarray}
where $I,J=L,R$.
This ratio can be served as a clear discriminant
to identify whether the wrong-sign operators are purely scalar type or not.

\section{Conclusion}

In the future $B$ experiments which can examine rare $B$ decays
with the branching ratio down to $10^{-10}$, it is worthwhile
to look for  the $\Delta S=-1$ and $+2$ processes which have extremely
small standard model background.  In this regards, we analyzed exclusive
two-body decay modes containing wrong-sign kaons in the final states,
signaling new physics effect.  As is well-known, the observation of
$B^- \to \pi^- K^{*0}$ or $K^- \Kb^{*0}$ provides a clean signal
for the existence of the wrong-sign amplitude.
However, with a reasonable efficiency in measuring $K_L$,
the wrong-sign amplitudes can be also probed
in  the $K_S$--$K_L$ asymmetry of
$\Bb \to \pi (K^{(*)}) K_{S,L}$ and $K_S K_L$, {\it etc},
or in the CP asymmetry of $B^\pm \to \pi^\pm (K^\pm) K_S$.
Combination of all the  observations will be useful
to investigate new physics beyond the standard model.
Thus, we consider it desirable to improve the  detection
efficiency for the identification of $K_L$ 
in the future $B$ experiments.
Observing  the wrong-sign kaons in the $B$ decays to one or 
two vector mesons
can lead us to study the origin of the wrong-sign amplitude.
For the $B$ decays driven by $\Delta S=+2$ transitions,
the type of the wrong-sign operators can  be identified
if anomalously high branching ratios are
measured for the modes $B^- \to K^- \Kb^{*0}, K_S K^{*-}$ and
$\Bb^0 \to K_S \Kb^{*0}$, or if  the observation of the modes,
$B^-\to K^- \Kb^{*0}$ and $\Bb^0 \to \Kb^{*0} \Kb^{*0}$, is made.

\medskip

{\bf Acknowledgment}: We would like thank Kazuo Abe,  Alex Kagan
and Pyungwon Ko for valuable discussions on the present and
related topics.
EJC was supported by  the Korean Research Foundation Grant,
KRF-2002-070-C00022.

\appendix
\section{Form factors}
The form factors for the $B$ decays used in our calculation
are defined as follows.

\newcommand{\wt}{\widetilde}
\newcommand{\imag}{\Im {\rm m}}
\newcommand{\real}{\Re {\rm e}}
\newcommand{\tanb}{\tan \! \beta}
\newcommand{\cotb}{\cot \! \beta}
\newcommand{\mto}{m^2_{\tilde{t}_1}}
\newcommand{\mttt}{m^2_{\tilde{t}_2}}
\newcommand{\mbo}{m^2_{\tilde{b}_1}}
\newcommand{\mbt}{m^2_{\tilde{b}_2}}
\newcommand{\ghat}{\hat{g}^2}
\newcommand{\htop}{\left| h_t \right|^2}
\newcommand{\hb}{\left| h_b \right|^2}

\noindent
$\bullet$  Meson decay amplitudes: 
\begin{eqnarray}
\langle \pi^-(q)|\bar{d}\gamma_\mu(1\pm\gamma_5) u|0\rangle &=&
\sqrt{2}\langle \pi^0(q)|\bar{u}\gamma_\mu(1\pm\gamma_5) u|0\rangle
\nonumber\\
&=&
-\sqrt{2}\langle \pi^0(q)|\bar{d}\gamma_\mu(1\pm\gamma_5) d|0\rangle
= \mp i f_\pi q_\mu\,, \nonumber \\
\langle K^-(q)|\bar{s}\gamma_\mu(1\pm\gamma_5) u|0\rangle &=&
-\langle K^0(q)|\bar{d}\gamma_\mu(1\pm\gamma_5) s|0\rangle
\nonumber\\
&=&
-\langle \bar{K}^0(q)|\bar{s}\gamma_\mu(1\pm\gamma_5) d|0\rangle
= \mp i f_K q_\mu \,,
\\
\langle K^{-*}(q,\epsilon)|\bar{s}\gamma_\mu u|0\rangle &=&
-\langle K^{0*}(q,\epsilon)|\bar{d}\gamma_\mu s|0\rangle
\nonumber\\
&=& -\langle \bar{K}^{0*}(q,\epsilon)|\bar{s}\gamma_\mu d|0\rangle =
f_{K^*}m_{K^*}\epsilon^*_\mu\,, \nonumber \\
\langle K^{-*}(q,\epsilon)|\bar{s}\sigma_{\mu\nu} u|0\rangle &=&
-\langle K^{0*}(q,\epsilon)|\bar{d}\sigma_{\mu\nu} s|0\rangle
\nonumber\\
&=&
-\langle \bar{K}^{0*}(q,\epsilon)|\bar{s}\sigma_{\mu\nu} d|0\rangle
= -i f^T_{K^*}(\epsilon^*_\mu q_\nu -\epsilon^*_\nu q_\mu)\,.
\end{eqnarray}
For the tensor form factors, it is useful to remember
$
\sigma_{\mu\nu}\gamma_5=
-\frac{i}{2}\epsilon_{\mu\nu\alpha\beta}\sigma^{\alpha\beta}
$ where $\epsilon_{0123}=+1$.

\noindent
$\bullet$ $B$ meson transition amplitudes:
\begin{eqnarray}
\langle \pi^-|\bar{d}\, \Gamma b|B^- \rangle &=&
\sqrt{2} \langle \pi^0|\bar{d}\,\Gamma b|\bar{B}^0 \rangle \,,
\nonumber \\
\langle K^-|\bar{s}\,\Gamma b|B^- \rangle &=&
\langle \bar{K}^0|\bar{s}\,\Gamma b|\bar{B}^0 \rangle  \,.
\end{eqnarray}

\fbox{Scalar currents}
\begin{eqnarray}
\langle M_1|\bar{q}q^\prime|M_2\rangle &=&
\frac{(p_{M_1}-p_{M_2})^\mu}{m_q-m_{q^\prime}}
\langle M_1|\bar{q}\gamma_\mu q^\prime|M_2\rangle \,,
\nonumber \\
\langle M_1|\bar{q}\gamma_5 q^\prime|M_2\rangle &=&
\frac{(p_{M_1}-p_{M_2})^\mu}{m_q+m_{q^\prime}}
\langle M_1|\bar{q}\gamma_\mu\gamma_5  q^\prime|M_2\rangle \,.
\end{eqnarray}

 \fbox{Vector currents}
\begin{eqnarray}
\langle
P(p)|\bar{q}\gamma_\mu(1\pm\gamma_5)b|B(p_B)
\rangle  &= &
\left[(p_B+p)_\mu-\frac{m_B^2-m_P^2}{q^2} q_\mu\right]F^{B\rightarrow P}_1(q^2)
\nonumber \\
&+&
\frac{m_B^2-m_P^2}{q^2} q_\mu F^{B\rightarrow P}_0(q^2) \,,
\end{eqnarray}
\begin{eqnarray}
\langle
V(p,\epsilon)|(V\pm A)_\mu|B(p_B)
\rangle  &= &
\pm i \epsilon^*_\mu(m_B+m_V)A_1^V(q^2)
\nonumber \\
&&
\mp i (p_B+p)_\mu ( \epsilon^*\cdot q)\frac{A_2^V(q^2)}{m_B+m_V}
\nonumber \\
&&
\mp i q_\mu (\epsilon^*\cdot q)
\frac{2m_V}{q^2}\left[A_3^V(q^2)-A_0^V(q^2)\right]
\nonumber \\
&&
+\epsilon_{\mu\nu\alpha\beta}\epsilon^{*\nu}q^\alpha p^\beta
\frac{2V^V(q^2)}{m_B+m_V} \,,
\end{eqnarray}
where $q=p_B-p$.
Here we have the relation;
$$
A_3^V(q^2)=\frac{m_B+m_V}{2m_V}A_1^V(q^2)
-\frac{m_B-m_V}{2m_V}A_2^V(q^2) \,.
$$

 \fbox{Tensor currents}
\begin{eqnarray}
\langle
P(p)|\bar{q}\sigma_{\mu\nu}q^\nu(1\pm\gamma_5)b|B(p_B)
\rangle  &= &
i\left[(p_B+p)_\mu q^2-q_\mu(m_B^2-m_P^2)\right]
\frac{F_2^{B\rightarrow P}(q^2)}{m_B+m_P} \,,
\end{eqnarray}
\begin{eqnarray}
\langle
V(p,\epsilon)|\bar{q}\sigma_{\mu\nu}q^\nu(1\pm\gamma_5)b|B(p_B)
\rangle  &= &
+i\epsilon_{\mu\nu\alpha\beta}\epsilon^{*\nu}q^\alpha p^\beta 2 T_1(q^2)
\nonumber \\
&&
\pm T_2(q^2)\left[\epsilon^*_\mu(m_B^2-m_V^2)-(\epsilon^*\cdot q)(p_B+p)_\mu
\right]
\nonumber \\
&&
\pm T_3(q^2) (\epsilon^*\cdot q)
\left[q_\mu-\frac{q^2}{m_B^2-m_V^2}(p_B+p)_\mu \right] \,.
\end{eqnarray}

For the form factors of the vector currents, we used the numerical
values taken in Ref.~\cite{JKK}.
For the tensor form factors,
we adopt the results of light cone sum rule calculation
\cite{ball,braun} given by
\begin{eqnarray}
&& f^T_{K^*} = 185 \mbox{ MeV} \,, \nonumber\\
&& F_2^{B\to \pi} = 0.296\,, \qquad F_2^{B\to K}=0.374 \,, \nonumber\\
&& T_1^{B\to K^*} = T_2^{B\to K^*} = 0.379\,,\quad
   T_3^{B\to K^*} = 0.260 \,.
\end{eqnarray}


\newpage
\begin{table}[ht]
\begin{center}
\vskip 0.5cm
\begin{tabular}{c|ccccc}
\hline
$w$ ($|w|$) & $C_{LL,RR}$ & $C_{LR,RL}$ & $C'_{LR,RL}$
     & $S_{LL,RR}$ & $S^\prime_{LL,RR}$ \\
\hline \hline $\pi^{0} K^{0}/\pi^{0}\bar{K}^{0}$ &
     $\pm0.75\pm0.017i$ & $\mp0.41\mp0.0092i$ & $\pm0.25\pm0.0056i$ &
     $\pm0.092\pm0.002i$  & $\mp0.037\mp0.0008i$ \\
$\pi^{-} K^{0}/\pi^{-} \bar{K}^{0}$ &
     $\pm0.65$ & $\mp0.36$ & $\pm0.22$  &
     $\pm0.079$ & $\mp0.032$ \\
$\pi^{0} K^{*0}/\pi^{0} \bar{K}^{*0}$ &
     $+1.7+0.064i$ & $+1.3+0.048i$ & $+0.43+0.016i$  &
     $-0.027-0.001i$ & $-0.082+0.003i$ \\
$\pi^{-} K^{*0}/\pi^{-} \bar{K}^{*0}$  &
     $+1.3$ & $+1.0$ & $+0.33$ & $-0.031$ & $-0.092$   \\
\hline \hline
$\bar{K}^{0}\bar{K}^{0}/\bar{K}^{0}{K}^{0}$ &
     $\mp1.2\pm0.50i$ & $\pm0.65\mp0.27i$ & $\mp0.42\pm0.18i$ &
     $\mp0.30\pm0.13i$ &  $\pm0.059\mp0.025i$   \\
${K}^{-}\bar{K}^{0}/{K}^{-}{K}^{0}$  &
     $\mp0.59\pm0.25i$ & $\pm0.32\mp0.14i$ & $\mp0.21\pm0.088i$ &
     $\mp0.15\pm0.063i$ & $\pm0.030\mp0.013i$ \\
$\bar{K}^{0}\bar{K}^{*0}/\bar{K}^{0}{K}^{*0}$  &
     $-2.2+0.91i$ & $-0.084+0.036i$ & $+0.43-0.18i$ &
     $+0.24-0.10i$ & $+0.028-0.012i$    \\
${K}^{-}\bar{K}^{*0}/{K}^{-}{K}^{*0}$ &
     $-1.2+0.53i$ & $-0.94+0.40i$ & $-0.31+0.13i$  &
     $+0.024-0.01i$ & $+0.071-0.03i$    \\
$\bar{K}^{*0}\bar{K}^0/\bar{K}^{*0}{K}^{0}$ &
     $+43-18i$ & $+1.7-0.72i$ & $-8.7+3.7i$  &
     $-4.8+2.0i$ & $-0.55+0.23i$     \\
${K}^{*-}\bar{K}^{0}/{K}^{*-}{K}^{0}$ &
     $+18-7.7i$ & $-17+7.3i$ & $-15+6.4i$  &  $-4.4+1.8i$ & $+0.87-0.37i$ \\
\hline
$\bar{K}^{*0}\bar{K}^{*0}/\bar{K}^{*0}{K}^{*0}$ &
     $\sqrt{2}(1+\xi)$ & $\sqrt{2}$ & $\sqrt{2}\xi$ &
     ${1.3 \xi}$ & ${1.3}$    \\
${K}^{*-}\bar{K}^{*0}/{K}^{*-}{K}^{*0}$ &
     $(1+\xi)$ & $1$ & $\xi$ &  $0.89\xi$ & $0.89$      \\
\hline
\end{tabular}
\end{center}
\caption[]{The amplitude ratios of the wrong-sign (WS) and the
right-sign standard model (SM)  processes; 
$w=A_{WS}/A_{SM}$
defined in Eqs.~(\ref{wbij1}) and (\ref{wbij2}). 
 The numerical values  are
 obtained with each non-vanishing  new physics coefficient
 $C^{(\prime)}_{IJ}$
 or $S^{(\prime)}_{II}$  shown in the first row,
 which is normalized as
 $C^{(\prime)dd},S^{(\prime)dd}=|V_{tb} V_{ts}^* (a_4-a_{10}/2)|
 =1.54\times10^{-3}$ or
 $C^{(\prime)ss},S^{(\prime)ss}=|V_{tb}V_{td}^*(a_4-a_{10}/2)|
 =2.91\times10^{-4}$.
In the last two lines, $|w|$ is  shown to denote the ratio
of $\sqrt{\Gamma_{WS}/\Gamma_{SM}}$ for each decay modes.
Here $\xi \equiv 1/N_c$ and $N_c=3$ is taken.
 }
\label{Table1}
\end{table}



\begin{thebibliography}{99}

\def\plb#1#2#3{Phys.\ Lett.\       {\bf B#1}  (#3) #2}
\def\npb#1#2#3{Nucl.\ Phys.\       {\bf B#1}  (#3) #2}
\def\prd#1#2#3{Phys.\ Rev.\        {\bf D#1}  (#3) #2}
\def\prl#1#2#3{Phys.\ Rev.\ Lett.\ {\bf #1}   (#3) #2}
\def\mpl#1#2#3{Mod.\ Phys.\ Lett.\ {\bf A#1}  (#3) #2}
\def\rep#1#2#3{Phys.\ Rep.\        {\bf #1}   (#3) #2}
\def\sci#1#2#3{Science             {\bf #1}   (#3) #2}
\def\astro#1#2#3{Astrophys.\ J.\   {\bf #1}   (#3) #2}
\def\epj#1#2#3{Eur.\ Phys.\ J.\   {\bf C#1}   (#3) #2}
\def\jhep#1#2#3{JHEP              {\bf #1}   (#3) #2}
\def\ptp#1#2#3{Prog.\ Theor.\ Phys.\ {\bf #1}  (#3) #2}

\bibitem{hlsz}
 K. Huitiu, C-D L\"u, P. Singer and D-X Zhang, \prl{81}{4313}{1998};
 \plb{445}{394}{1999};
 S. Fajfer and P. Singer, \prd{62}{117702}{2000}.
\bibitem{superB}
 See, for instance, the documents at
 fermi.phy.uc.edu/HyperNews/get/forums/SuperBaBar.html.
\bibitem{tevb}
 K. Anikeev, {\it et al.}, ``B Physics at the Tevatron: Run II and Beyond'',
 hep-ph/0201071;
 BTeV Collaboration, A. Kulyavtsev  {\it et al.}, hep-ph/9809557.
\bibitem{LHCb}
   J.\ Baines {\it et al.}, ``$B$ decays at the LHC'', hep-ph/0003238.
\bibitem{bhatt}
 G. Bhattacharyya, hep-ph/9709395 
\bibitem{ali}
 A.~Ali, G.~Kramer and C-D L\"u, \prd{58}{094009}{1998};
            \prd{59}{014005}{1999}.
\bibitem{JKK} J.-H. Jang, Y.G. Kim, and P. Ko, \prd{59}{034025}{1999}.
\bibitem{baba1}
 BaBar Collaboration, B. Aubert {\it et al.}, hep-ex/0304006.
\bibitem{baba2}
 BaBar Collaboration, B. Aubert {\it et al.}, talk presented at the
 XVIIth Rencontres de la Vallee d'Aoste, hep-ex/0303022.
\bibitem{tomura}
 M. Bona, talk presented in FPCP03, Paris, June 2003;
 T. Tomura, hep-ex/0305036.
\bibitem{chen}
 Y-H.~Chen, H-Y.~Cheng, B.~Tseng and K-C.~Yang, \prd{60}{094014}{1999}
\bibitem{bbns}
M. Beneke, G. Buchalla, M. Neubert, C.T. Sachrajda,
\prl{83}{1914}{1999}; \npb{606}{245}{2001}.
\bibitem{du}
 D.~Du, H.~Gong, J.~Sun, D.~Yang and G.~Zhu,
\prd{65}{074001}{2002}.
\bibitem{keum}
 Y.-Y.~Keum, H.-n. Li and A.I. Sanda,
\plb{504}{6}{2001};
 \prd{63}{054008}{2001}.


\bibitem{bigi}
I.I.~Bigi and H.~Yamamoto, \plb{349}{363}{1995}.

\bibitem{trojan}
 Y. Grossman, M. Neubert and A.L. Kagan,
 JHEP {\bf 9910} (1999) 029.

\bibitem{yoshi}
 T. Yoshikawa, hep-ph/0306147.

\bibitem{alex}
A. Kagan, private communication.

\bibitem{ball} P. Ball, JHEP {\bf 09} (1998) 005.
\bibitem{braun} P. Ball and V.M. Braun, \prd{58}{094016}{1998}.



\end{thebibliography}
\end{document}